\documentclass[twocolumn,showpacs,amsmath,amssymb,prd,nofootinbib]{revtex4}
\usepackage{graphicx}
\usepackage{epsfig}
\usepackage{enumerate}
\def\beq{\begin{equation}}
\def\eeqno#1{\label{#1}\end{equation}}

\def\rarrow{\rightarrow }

\def\dleft{\rlap{{\it D}}\raise 8pt
\hbox{$\scriptscriptstyle\Leftarrow$}}
\def\dright{\rlap{{\it
D}}\raise 8pt\hbox{$\scriptscriptstyle\Rightarrow$}}

\def\kms{{\rm km~s^{-1}}}
\def\cmss{{\rm cm~s^{-2}}}
\def\pc{~{\rm pc}}
\def\kpc{~{\rm Kpc}}

\def\msun{M_{\odot}}
\def\az{a_{0}}

\def\l0{\ell_{0}}

\def\l{\lambda}

\def\m{\mu}
\def\n{\nu}

\def\xlimin{{x\rarrow\infty \atop{\raise 1pt\hbox to 30pt
{\rightarrowfill}}}}
\def\limlim#1#2{{#1\rarrow #2 \atop{\raise 1pt\hbox to 30pt
{\rightarrowfill}}}}

\def\S{\Sigma}

\def\gN{g\_N}

\def\m{\mu}

\def\n{\nu}

\def\_#1{_{\scriptscriptstyle #1}}
\def\^#1{^{\scriptscriptstyle #1}}

\begin{document}
\title{MOND and the dynamics of NGC 628}
\author{Mordehai Milgrom}
\affiliation{Department of Particle Physics and Astrophysics, Weizmann Institute}

\begin{abstract}

Aniyan et al. (2018) have recently published direct measurements of the baryonic mass distribution and the rotation curve of the almost-face-on disc galaxy NGC 628. While its very low inclination makes this galaxy anything but ideal for rotation-curve analysis, these new results, taken at face value, have interesting ramifications for MOND.  The methods employed afford a direct determination of the stellar mass in the disc, which, in turn, affords a {\it parameter-free} MOND prediction of the rotation curve, which I show. In comparison, the dark-matter fits that Aniyan et al. present have two free parameters. To boot, these results further negate an earlier claim deleterious to MOND. It is that stellar  $M/L$ ratios deduced from vertical velocity dispersions in disc galaxies are rather lower than what is required by MOND fits to rotation curves. Specifically, it was claimed that even high-surface-density discs are, by and large, sub maximal; viz., that they show substantial mass discrepancies near their center. This is contrary to the prediction of MOND that in such high-acceleration regions only small discrepancies should appear, if any to speak of. Such claims of low $M/L$ values have been rebutted before, and the fallacy that may have led to them pointed out. The new results strongly buttress these rebuttals.
\end{abstract}
\maketitle
\section{Introduction}
MOND predicts the dynamics of galaxies -- in particular, the rotation curves (RCs) of disc galaxies -- given only the baryon mass distribution in them.\footnote{For recent reviews of MOND see Famaey and McGaugh (2012) and Milgrom (2014).} For some galaxies the baryons are dominated by cool gas whose mass can be determined directly, and then an almost-parameter-free MOND prediction of the RC can be made. In most cases, however, stellar components contribute significantly, and it is necessary to convert starlight to mass, a conversion that is encapsuled in the $M/L$ conversion factor. Population-synthesis models give us some idea of what this parameter should be for various galaxies. But these are more statistical in nature -- i.e., not meant to match individual galaxies -- and are not accurate enough to the level required to match the accuracy of the MOND predictions to the quality of the observed RCs.
\par
If it were possible to measure directly the stellar mass distribution, or at least the average $M/L$ value for individual galaxies it would be a qualitative jump in the power of MOND, as it would allow parameter-free predictions of the dynamics (assuming that other galaxy attributes that go into the predictions, such as the distance, are measured accurately enough).
\par
Unlike the case for MOND, the dark-matter paradigm does not, in any case, predict galaxy dynamics from the baryon mass distribution. It only fits the observations with a dark-matter halo, itself parametrized by several free parameters. So not knowing $M/L$ just adds another fit parameter, which is not such a qualitative hinderance.
\par
One way in which it was proposed to measure stellar-disc masses directly is to combine measurements of the vertical velocity dispersions in the disc with those of the scale height, and determine the surface mass density assuming vertical dynamical equilibrium. This method has important limitations. For example it is difficult to measure for the same galaxy both the scale height (which requires nearly edge on orientation) and the velocity dispersions (which requires nearly face-on ones).
\par
The DiskMass project (Bershady 2010a,b) was dedicated to this end. It measured the vertical velocity distributions of many low-inclination disc galaxies, and combined them with estimated scale heights based on empirical relations found in other studies of high-inclination galaxies.
\par
The surprising result of the DiskMass analysis was that the $M/L$ values found were rather low. In particular, they were about a factor of two lower than results of population-synthesis estimates, and of other, indirect, estimates. Moreover, Angus, Gentile and Famaey (2016) showed that due to some omission, these DiskMass estimates of $M/L$ should have been even lower -- typically $\sim 1/3$ of the population-synthesis and other estimates.
\par
Such low disc masses and $M/L$ ratios imply that discs are by and large `sub maximal'; i.e., the baryonic rotation curve falls substantially below the observed RC in the inner parts, even for high-surface brightness galaxies.
\par
These findings run against what had been the common thought, and,
if true, would have particular significance in the context of MOND. In particular, MOND {\it predicts} that high-surface-brightness galaxies, whose RCs, $V(r)$, show high accelerations in the inner parts [$g=V^2(r)/r\gg \az$, where $\az$ is the MOND acceleration constant] should show at most small mass discrepancies near the center, unlike the substantial discrepancies the DiskMass claims pointed to. 
\par
This immediately apparent clash has found a more quantitative expression in the MOND analysis of Angus et al. (2015). Their analysis was detailed, but in essence what they found was that for the individual galaxies of the DiskMass sample, the best $M/L$ values they deduced from MOND analysis of the RCs -- which do agree with population-synthesis estimates -- were larger by a factor of $\sim 2$ than those implied by the DiskMass analysis.
\par
In Milgrom (2015), I criticized these conclusions on various grounds of conflict with other estimates of the $M/L$ values, suggesting that the DiskMass velocity dispersions, while perhaps correctly measured, are wrongly interpreted in the analysis. My conjectured culprit was that DiskMass were combining dispersions and scale heights that do not belong to the same stellar population used as test particles in the analysis: The dispersions are heavily weighted by young disc stars whose scale heights and dispersions are small, while the scale heights used in the analysis are dominated by older populations of higher dispersions and higher scale heights. One should, of course use both parameters for the same population.\footnote{Bosma (1999) already pointed such a mismatch as a potential source of erroneous deductions.}
\par
I found that if the DiskMass velocity dispersions are only $\sim 30\%$ smaller than the ones that represent the stellar populations whose scale height is used, the analysis of Angus et al. (2015) would yield very good agreement with the MOND predictions.
\par
At the same time, Aniyan et al. (2016) -- who have set on a project to follow the procedure with due care to distinguishing the different populations -- have criticized the DiskMass claims on the same grounds, substantiating their criticism by an analysis of vertical dynamics near the sun in the Milky Way. They found that indeed, doing things correctly gives an $M/L$ value about twice as large as that gotten by following the DiskMass method of using integrated (over the stellar populations) velocity dispersions. This corresponds, indeed, to an artificial  underestimate of $\sim 30\%$ in the dispersions.
\par
Still, the possibility of such a fallacy in the deductions of DiskMass had remained moot.
\par
Now, Aniyan et al. (2018) present measurements and analysis of the vertical velocity dispersions in the almost-face-on galaxy NGC 628 (aka M74, aka UGC 1149), being careful to account for the separate contributions of young and old populations. They also present a measurement of the RC of this galaxy.
\par
In many regards this is just another galactic RC. If anything, its very small inclination ($\sim 8\^o$) renders it a rather dubious case to analyze. Exacting past RC analyses of MOND predictions have excluded such a low-$i$ galaxy from the outset.
However, in light of what I said above, the results of Aniyan et al. (2018) do have some important implications for MOND. It is the purpose of this short note to limelight these.
\par
In particular, the new analysis demonstrates clearly how adopting the DiskMass procedure can underestimates the appropriate velocity dispersion by about $30\%$. Aniyan et al. (2018) give for NGC 628 an $M/L$ value, in the $3.6 \m$ band, of $0.75\pm 0.18$ solar units, consistent with stellar population values ($\sim 0.6$ solar units), and with typical MOND best fit values,\footnote{Angus et al. 2015 find for the sample of DiskMass galaxies they studied MOND best-fit values in the K band mean value of of $.55\pm.15$} but a factor of $\sim 2$ larger than the typical ones claimed by DiskMass.
\par
The same is the case for other two galaxies: According to the abstract of Aniyan's thesis: ``We then use a sample of three nearby, relatively face-on spirals (NGC 628, NGC 6946, and NGC 5457) to extract the vertical velocity dispersion of the hot thin disc. ... In all three galaxies, using these two kinematic tracers, we were able to extract the velocity dispersion of the hot component, which we then used along with the scale height (for the same component) to determine the surface mass densities. The central surface mass densities that we derive are typically at least a factor of 2 higher than previous studies."
\par
Note that the $z$-dynamics analysis of Aniyan et al. (2018) is based on Newtonian dynamics. Generally, these dynamics have to be treated according to MOND, which would yield smaller dynamical surface densities. However, In the inner parts ($R\lesssim 1.5\kpc$) the MOND correction is small; so the central surface density and M/L ratio gotten by Aniyan et al. are adequate by MOND as well. These alone enter the analysis below.

\section{The MOND predicted rotation curve}
Given the Newtonian rotation curve of a disc galaxy, $V\_N(r)$, derived from its baryonic mass distribution, the parameter-free, predicted MOND RC is gotten from the relation (Milgrom 1983)\footnote{This holds exactly in `modified-inertia' formulations, with small departures for `modified-gravity' ones.}
\beq g=\gN\n(\gN/\az), \eeqno{mala}
with the Newtonian acceleration $\gN=V\_N^2(r)/r$.
\par
To calculate the MOND RC for NGC 628, I used $V\_N(r)$ given in Figure 15 of Aniyan et al. (2018) in eq. (\ref{mala}), with the standard $\az=1.2\times 10^{-8}\cmss$, and $\nu(y)=(1-e^{-y^{1/2}})^{-1}$ [introduced, in Milgrom and Sanders (2008) as part of a MOND analysis of cluster lensing, and used
repeatedly for MOND analysis of rotation curves, e.g. in McGaugh (2008) and in Famaey and McGaugh (2012), and for analyzing x-ray ellipticals in Milgrom (2012)].
\par
The baryonic RC given by Aniyan et al. is modeled by that of an exponential disc with a scale length of $\approx 3.9 \kpc$ and a central surface density $\S(0)=505\pm 170\msun/\pc^2$ (and an adopted scale height of $\approx 400\pc$, to which the RC is insensitive). This model disc includes the contributions of the stellar and gas discs. To this is added a bulge component with some assumed, model $M/L$, which unlike $M/L$ for the disc, is not measured. The value of $\S(0)$ is the crucial parameter and is taken from their $z$-dynamics at small radii, which, as alluded to above, is correct in MOND as well because the accelerations in the inner parts are high. For example, at $0.5$ and $1\kpc$, $V^2/r\az\approx 6.8$ and $2.5$ respectively.

\begin{figure}
\begin{center}
\includegraphics[width=1.0\columnwidth]{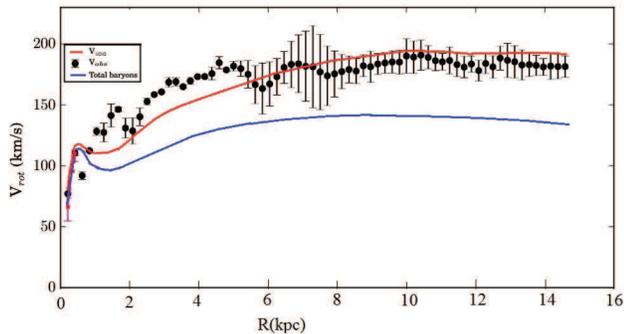}
\caption{The predicted (zero-parameter) MOND rotation curve (RC) of NGC 628  (red line), calculated from eq. (\ref{mala}), based on the baryonic RC of Aniyan et al. (2018) (blue line). The measured RC (assuming a constant inclination of $8\^o$) is shown as the points with error bars.}\label{fig1}
\end{center}
\end{figure}

The MOND curve is shown in Figure \ref{fig1}, together with $V\_N(r)$ and the measured RC, also from Fig. 15 of Aniyan et al. (2018).
\par
The results of Aniyan et al. are still subject to substantial uncertainties. Fore example, the inclination was fixed in the derivation of the actual RC to $i=8\^o$, but even a few degrees uncertainty in $i$  would produce large uncertainties in the deduced RC. Also, in the region  $2-6\kpc$ where the deviation of the MOND prediction is more apparent, the measured rotation curves for the receding and approaching sides of the galaxy are rather different (a difference of $\sim 50\kms$; see Fig. 13 of Aniyan et al.). This is not reflected in the error bars shown. In fact, the MOND curves follows very closely the receding half of the measured RC in this region.

\end{document}